\documentclass[fleqn,usenatbib]{mnras}
\usepackage{graphicx}
\usepackage{float}
\usepackage{wrapfig}
\usepackage{amsmath}
\usepackage{amssymb}

\usepackage{bm}
\usepackage[font=small,labelfont=bf]{caption}
\usepackage{citesort}
\usepackage{threeparttable}
\usepackage{comment}
\usepackage[dvipsnames]{xcolor}

\usepackage[normalem]{ulem}  

\usepackage[textwidth=1in]{todonotes}
\usepackage{multicol, cuted}

\title[HD/H$_2$ ratio in the diffuse ISM]{HD/H$_2$ ratio in the diffuse interstellar medium}

\author[S. A. Balashev and D. N. Kosenko]{
S. A. Balashev,$^{1}$\thanks{E-mail: s.balashev@gmail.com}
D. N. Kosenko,$^{1}$\thanks{E-mail: kosenkodn@yandex.ru}
\\
$^{1}$Ioffe Institute, 26 Politeknicheskaya st., St.\ Petersburg, 194021, Russia\\
}

\date{Accepted 2019 November 27. Received 2019 November 26; in original form 2019 September 18}

\pubyear{2019}

\begin{document}

\label{firstpage}
\pagerange{\pageref{firstpage}--\pageref{lastpage}}
\maketitle

\begin{abstract}

We present a semi-analytical description of the relative HD/H$_2$ abundance in the diffuse interstellar medium. We found three asymptotics of the relative HD/H$_2$ abundance for different parts of the medium and their dependence on the physical parameters, namely, number density, intensity of the ultraviolet field, cosmic ray ionization rate and metallicity. Our calculations are in a good agreement with the full network calculations using Meudon PDR code. We found that in the case of low metallicity and/or higher cosmic ray ionization rate, HD formation rate is significantly enhanced, HD/H$_2$ ratio increases, and the D\,{\sc i}/HD transition occurs at a lower penetration depth of UV radiation than the H\,{\sc i}/H$_2$ transition. This can explain the observed difference in the HD/H$_2$ abundance between the local and high-redshift measurements.

\end{abstract} 


\section{Introduction}


The HD molecule is one of the most abundant molecule in the cold interstellar medium (ISM). Since HD is predominantly formed in ion-neutral reaction involved $\rm H_2$, it can be used as a tracer of $\rm H_2$.
However, the numerical modelling of ISM indicates that the relative HD/H$_2$ abundance is very sensitive to physical conditions in the medium \citep[e.g.][]{LePetit2002}. Previous studies discussed that the HD/H$_2$ abundance is sensitive to the cosmic ray ionization rate and number density \citep[see e.g.][]{Hartquist1978, Liszt2003, Cirkovic2006, Liszt2015}.
However, its dependence on the main physical parameters is not extensively studied and is still poorly constrained observationally.

The only available to date method to directly measure HD/H$_2$ relative abundance is absorption line spectroscopy of the electronic transitions. Observations with UV space telescopes ({\it Copernicus} and {\it FUSE}) found that $N_{\rm HD}/2N_{\rm H_2}$ (where $N$ is the column density in cm$^{-2}$) measured towards bright stars in our Galaxy is significantly below the D/H isotopic ratio \citep{Lacour2005, Snow2008}. 
HD and H$_2$ absorption lines are also detected in the spectra of high-z quasars, to be associated with so-called Damped Lyman alpha systems \citep[][DLA]{Wolfe2005}. Measurements of $N_{\rm HD}/2N_{\rm H_2}$ in DLAs are particularly valuable, since they typically probe a wide range of physical conditions. The problem is that the cold ($\sim 100$\,K) and dense (with number densities, $n \gtrsim 100$\,cm$^{-3}$) phase of ISM probed by H$_2$/HD-bearing components in DLAs has a small cross section, and therefore the fraction of H$_2$/HD-bearing DLAs is small, $\sim4$ per cent \citep{Balashev2018}. Therefore, since the first detection of HD at high redshifts \citep{Varshalovich2001}, only 11 H$_2$/HD-bearing DLAs have been detected so far  
\citep[at $z>2$,][]{Noterdaeme2008, Noterdaeme2010, Ivanchik2010, Balashev2010, Tumlinson2010, Klimenko2015, Ivanchik2015, Klimenko2016, Balashev2017, Noterdaeme2017, Kosenko2018, Rawlins2018}.
Already available measurements indicate that $N_{\rm HD}/2N_{\rm H_2}$ at high redshifts is typically higher than those in our Galaxy \citep{Balashev2010, Tumlinson2010}. 
However, the expected astration of deuterium is small \citep{Dvorkin2016} and cannot solely explain the observed discrepancy. 

In this paper we present a simple semi-analytical description of the relative HD/H$_2$ abundance in diffuse ISM. We found that it is strongly varied with the physical conditions (metallicity, cosmic ray ionization rate, UV field strength and number density) in diffuse ISM. For a wide range of the physical conditions, we found that our calculations well agree with the calculations using the Meudon PDR code, which solves the full radiative transfer and chemical equation network. However, our calculations in presented formalism are much faster than $\textit{Meudon PDR}$ code, and therefore, they can be efficiently used to study the parameter space to constrain physical conditions from the observed $N_{\rm HD}/2N_{\rm H_2}$ abundance. We also show that HD production at low metallicity is significantly enhanced and therefore the observed discrepancy between the high-z DLAs and Milky-Way measurements of $N_{\rm HD}/2N_{\rm H_2}$ can be naturally explained. 

\section{Semi-analytical description}

We consider the homogeneous medium with the total hydrogen number density $n_{\rm H}^{\rm tot}$ (in the following we will use $n_2=n_{\rm H}^{\rm tot} / 100\,\rm cm^{-3}$) and   metallicity $Z$ (relative to solar) exposed by the UV field of strength $\chi$ \citep[in the units of Draine field, ][]{Draine1978} and cosmic rays (with primarily ionization rate per hydrogen atom, $\zeta_{\rm p}$, measured in units $10^{-17}$\,s$^{-1}$). All reaction rates in the following have little temperature dependence at characteristic values of the temperature in the diffuse ISM \citep[50-200\,K, see][]{Balashev2017}, therefore we fixed temperature at $T=100$\,K.

In the diffuse ISM, the equilibrium number density of HD molecules, $n_{\rm HD}$, is determined by a balance between the formation and destruction processes. Two main channels of the HD formation are the gas-phase reaction
\begin{equation}
\label{eq:H2+D+}
\rm H_2 + D^{+} \to \rm HD + H^{+}
\end{equation}
and the formation of HD from atomic D on the surface of dust grains. The main channel of the HD destruction is the photodestruction\footnote{The chemical destruction related to the reverse reaction of \eqref{eq:H2+D+} plays the role only in the central very selfshielded part the medium, where the almost all D in HD and H in H$_2$. Therefore we neglect this channel.} associated with UV pumping to the excited electronic states in resonant lines of HD Lyman and Werner bands. A fraction \citep[$\sim 15$ per cents, see][]{LePetit2002} of excited HD molecules goes back to the continuum of the ground electronic state, i.e. they dissociate. When the UV radiation penetrates into the medium, HD lines in which excitation occurs become saturated and the UV pumping rate decreases\footnote{There also can be an additional absorption of the UV field by dust, H$_2$ and H\,I lines}, and, therefore, the photodestruction rate reduces. This is a well-known self-shielding mechanism that regulates \ion{D}{I}/HD (and \ion{H}{i}/H$_2$) transitions in the ISM. 
It is usually specified by a self-shielding function $S^{\rm HD} (N_{\rm HD}, N_{\rm H_2})$ \citep{Wolcott_Green2011}, which shows how the photodestruction rate decreases as a function of the HD and H$_2$ column densities \citep{Wolcott_Green2011} and a Doppler parameter, $b$ (in the following we set $b=2$\,km/s),
relative to an unattenuated photodestruction rate $\chi D^{\rm HD}$. Based on PDR Meudon calculations (see Section~\ref{sect:Meudon}) we adopt $D^{\rm HD} = 3.2\times 10^{-11}$\,s$^{-1}$ for the medium irradiated uniformly by the UV field with the Draine shape. 

Therefore under the steady state assumption and plane-parallel geometry, when one side of the medium is exposed by an unattenuated UV field, we can write
\begin{equation}
\label{eq:HD}
F^{\rm HD} n_{\rm H_2} n_{\rm D^{+}} + R^{\rm HD} n^{\rm tot}_{\rm H} n_{\rm D} = \frac12 \chi D^{\rm HD} S^{\rm HD} e^{-\tau_g}n_{\rm HD}, 
\end{equation}
where $F^{\rm HD}$ ($\approx2 \times 10^{-9}$\,cm$^{3}$s$^{-1}$, \citealt{LePetit2002}) is the HD chemical formation rate in reaction~\eqref{eq:H2+D+}, $R^{\rm HD} \approx 6.3\times 10^{-17} Z$ cm$^{3}$s$^{-1}$ is the HD formation rate on the dust grains \citep[][]{LePetit2002}, $\tau_g$ is the optical depth attributed with attenuation of the UV field by dust. The latter can be expressed as a function of a total hydrogen column density: $\tau_g = \sigma_g (N_{\rm H} + 2 N_{\rm H_2})$, where $\sigma_g = 1.9 \times 10^{21} Z$\,cm$^{-2}$ \citep[][]{Draine2003} is the dust grain LW-photon absorption cross section per one hydrogen nucleon \citep{Sternberg2014}.

In diffuse ISM, the D$^{+}$ abundance is mainly determined by the charge-exchange reaction
\begin{equation}
\label{eq:D_H_exchange}
{\rm H^{+}} + {\rm D} \longleftrightarrow {\rm D^{+}} + {\rm H}    
\end{equation}
and a D$^{+}$ destruction due to the reaction~\eqref{eq:H2+D+}. The reverse reaction to \eqref{eq:H2+D+} is significantly suppressed by endothermicity of $\sim460$\,K in the cold gas. Therefore, we get
\begin{equation}
\label{eq:n_DII}
n_{\rm D^{+}} = \dfrac{k}{k^{\prime}}\cfrac{n_{\rm H^{+}}n_{\rm D}}{n^{\rm tot}_{\rm H} + 2 n_{\rm H_2} B } \approx \dfrac{k}{k^{\prime}}\cfrac{n_{\rm H^{+}}n_{\rm D}}{n^{\rm tot}_{\rm H}},
\end{equation}
where $k$ and $k^{\prime}$ are rates of direct and reverse reaction \eqref{eq:D_H_exchange}, respectively, $n^{\rm tot}_{\rm H} = n_{\rm H} + 2 n_{\rm H_2}$ is the hydrogen number density, $B = F^{\rm HD} / 2 k^{\prime} - 1$ (following the values of the rates given by \citealt{LePetit2002}, $F^{\rm HD} \approx 2 k^{\prime}$ and, hence, $B\approx 0$). 

\begin{figure*}
    \begin{minipage}{0.5\textwidth}
        \includegraphics[width=1.0\textwidth]{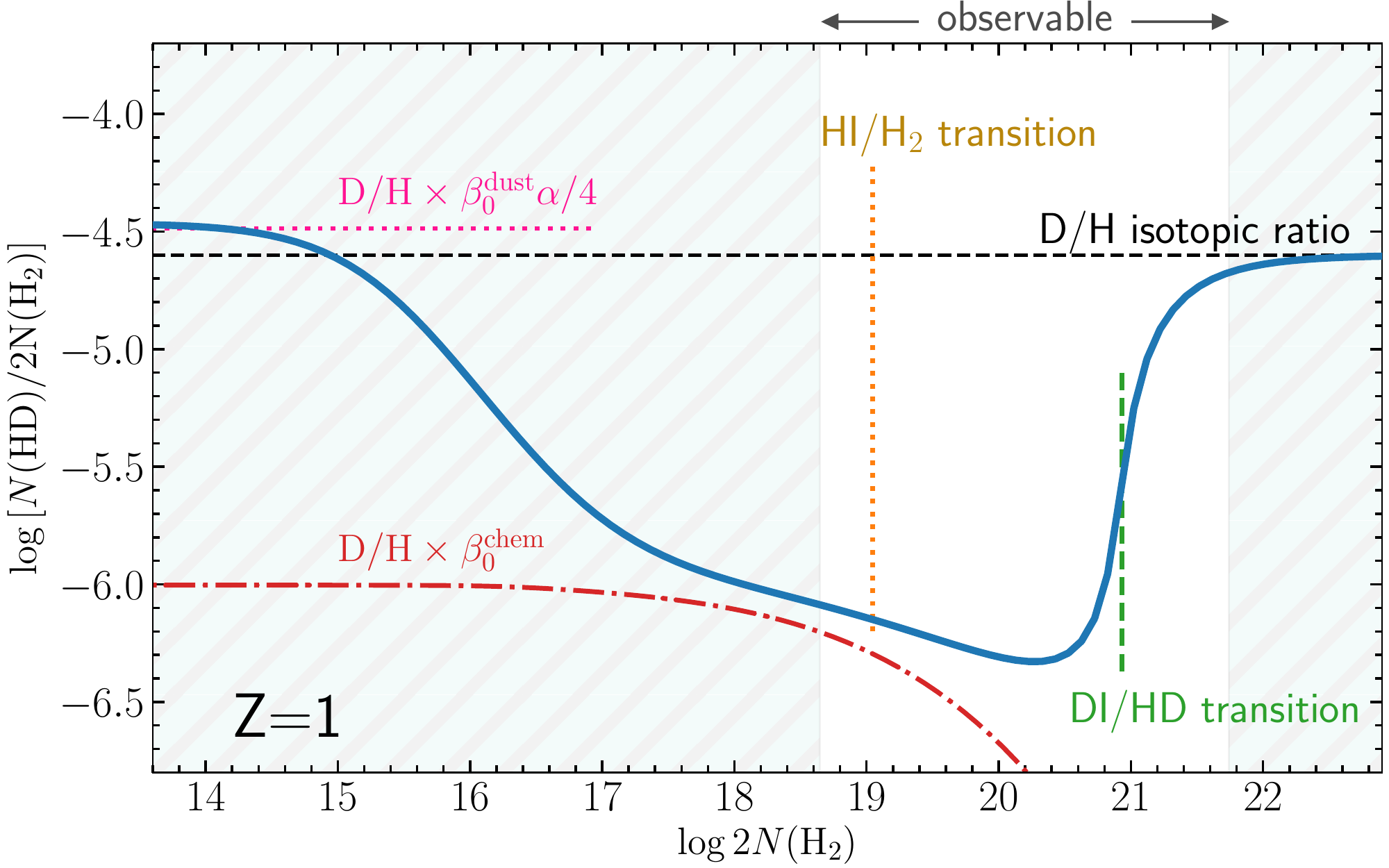}
    \end{minipage}%
    \begin{minipage}{0.5\textwidth}
        \includegraphics[width=1.0\textwidth]{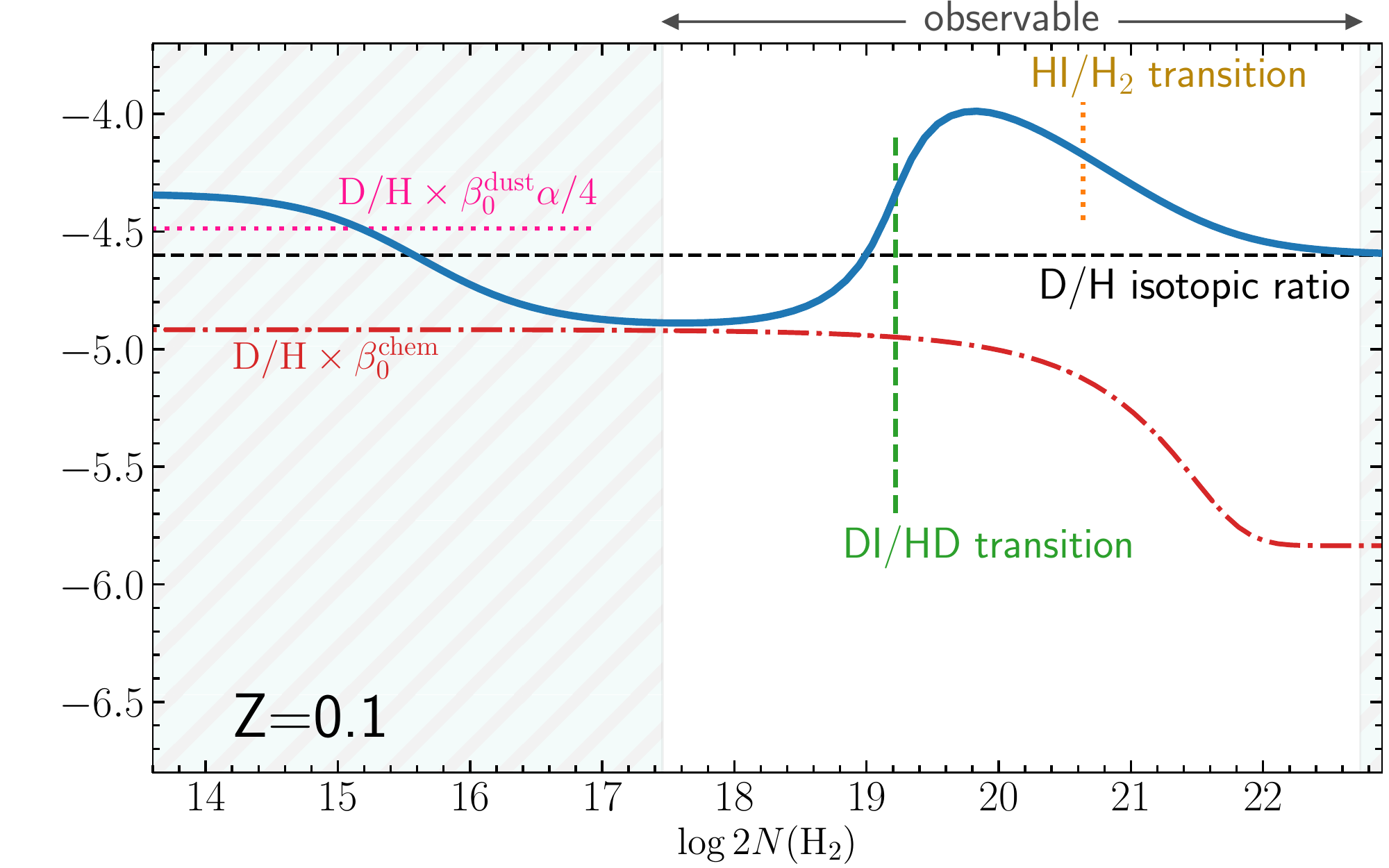}
    \end{minipage}
    \hspace{0.3cm}
    \caption{$\log N(\rm HD) / N(\rm H_2)$ as a function of $\log N(\rm H_2)$ calculated for $n_2=1$, $T=100$\,K, $\chi=1$ and  $\zeta_{\rm p}=3$, but for values of $Z=1$ and $Z=0.1$ for left and right panels, respectively. The solid blue curve is solution of equation~\eqref{eq:HD_H2_local}, the magenta dotted line represents the asymptotic on the edge of the cloud \eqref{eq:HD_H2_local_dust}, the red dashed line is the asymptotic in case of unattenuated field neglecting the formation of HD on dust \eqref{eq:HD_H2_local_chem} and the black dashed line corresponds to the D/H isotopic ratio. The green dashed and orange dotted vertical lines indicate H$_2$ column density of D\,{\sc i}/HD \eqref{eq:D_HD_tr_point} and H\,{\sc i}/H$_2$ transition \citep{Sternberg2016}, respectively. The region of $\log N(\rm HD) / N(\rm H_2)$ that is observable with a current instrumentation is caclulated using $\log N_{\rm HD} > 12.5$ and $A_V < 3$.}
    \label{fig:main}
\end{figure*}

The number density of protons, $n_{\rm H^+}$, can be derived from the ionization balance taking into account that it is mainly produced by the cosmic rays ionization of H and H$_2$ and destructed by a radiative recombination (with a rate $\alpha^{rr}\approx 1.1 \times 10^{-11}$\,s$^{-1}$, \citealt{LePetit2002})
 and a neutralization by grains (with a rate $\tilde\alpha^{gr}\approx  1.6 \times 10^{-10} \chi^{-1} n_2 Z$\,s$^{-1}$,
\citealt{Weingartner2001}):
\begin{equation}
    \label{eq:CR_H}
    k_{\zeta}^{\rm H} n_{\rm H} + k_{\zeta}^{\rm H_2} n_{\rm H_2} = \alpha^{rr} n_{\rm H^{+}} n_e + \tilde\alpha^{gr} n_{e} n_{\rm H^{+}} 
\end{equation}
where $k_{\zeta}^{\rm H} = \zeta_{\rm p} (1 + \phi_s) 10^{-17}$\,s$^{-1}$
is a cosmic ray ionization rate of atomic hydrogen (the factor $\phi_{\rm s}\approx0.67$ takes into account the secondary ionization, \citealt{Draine2011}), and $k_{\zeta}^{\rm H_2} = 0.1k_{\zeta}^{\rm H}$
is the rate of dissociation of H$_2$ by cosmic rays, resulting in direct H$^{+}$ production \citep[see][]{LePetit2002}.  There are additional channels of H$^+$ formation and destruction that can be taken into account in equation~\ref{eq:CR_H}. However, we found that $\rm H^+$ production by $\rm D^+$ and destruction by $\rm D$ are precisely balance each other. Also production of $\rm H^+$ through $\rm He^+$, $\rm H_2^+$ and charge exchange reactions with $\rm O$ also almost compensate each other (resulting in $\sim 20$ per cent corrections in the worst cases) for the ranges of the physical conditions concerned in this paper. Therefore for simplicity we did not take into account aforementioned reactions. 
The number density of electrons in the cold ISM is $n_{e} = n_{\rm H^+} + n_{\rm C^+}$\footnote{The fraction of electrons that comes other species, e.g. He$^{+}$, O$^{+}$, H$_2^+$ and H$_3^+$, is negligible in the cold ISM.}. 
We assume that C$^{+}$ is the dominant ionization state of carbon  with gas phase abundance $n_{\rm C}/n^{\rm tot}_{\rm H} \equiv x_{\rm C} = 2.7\times 10^{-4}Z d$, where $d$ is the depletion of the carbon, which is $\approx0.5$ at Solar metallicity and $\approx 1$ at sub-Solar metallicities. The solar undepleted abundance of $\rm C$ is taken from \citealt{Asplund2009}.
For reasonable ranges of physical parameters 
an approximate solution of equation~\eqref{eq:CR_H} gives

\begin{equation}
    \label{eq:f_HII}
    f_{\rm H^{+}} \equiv \frac{n_{\rm H^+}}{n^{\rm tot}_{\rm H}} = \frac{x_{\rm C}}{2}\left(\sqrt{\frac{4(k^{\rm H}_{\zeta} - f_{\rm H_2}(k^{\rm H}_{\zeta} - 0.5k^{\rm H_2}_{\zeta}))} {(\tilde\alpha^{gr} + \alpha^{rr}) n^{\rm tot}_{\rm H} x_{\rm C}^2} + 1} -1\right), 
\end{equation}
where $f_{\rm H_2} = \frac{2n_{\rm H_2}}{\left(n_{\rm H} + 2n_{\rm H_2}\right)}$ is H$_2$ molecular fraction.

Substituting equation~\eqref{eq:n_DII} and \eqref{eq:f_HII} into equation~\eqref{eq:HD} we obtain 
\begin{equation}
    \label{eq_HD_2}
    n_{\rm HD} = n_{\rm D} \left(\beta_{\rm chem} f_{\rm H_2} + \beta_{\rm dust} \right),
\end{equation}
where we introduce
\begin{align}
    \label{beta}
    \beta^{\rm chem} &= \frac{k F^{\rm HD} n^{\rm tot}_{\rm H} f_{\rm H^+}} {k^{\prime}\chi D^{\rm HD} S^{\rm HD} e^{-\tau_g}} \equiv \frac{\beta^{\rm chem}_0}{S^{\rm HD} e^{-\tau_g}}  \\
    \beta^{\rm dust} &= \frac{2 R^{\rm HD} n^{\rm tot}_{\rm H}}{\chi D^{\rm HD} S^{\rm HD} e^{-\tau_g}} \equiv \frac{\beta^{\rm dust}_0}{S^{\rm HD} e^{-\tau_g}},
\end{align}
where $\beta_0^{\rm chem}$ and $\beta_0^{\rm dust}$ are the values of $\beta^{\rm chem}$ and $\beta^{\rm dust}$ in the case of the unattenuated UV field, i.e. $S_{\rm HD} = 1$ and $\tau_g = 0$, which can be written as functions of the physical conditions:
\begin{align}
    \label{beta_0}
    \beta_0^{\rm chem} & = 0.6\dfrac{n_2Zd}{\chi}\left(\sqrt{\dfrac{0.8\zeta_{\rm p}(1 - 0.95f_{\rm H_2})}{n_2Z^2d^2(1 + 15\chi^{-1}n_2Z)} + 1 } - 1\right)  \\
    \beta^{\rm dust}_0 & = 1.2\times 10^{-4}  \cfrac{n_{2} Z}{\chi}.
\end{align}

Taking into account that $n_{\rm D} + n_{\rm HD} = n^{\rm tot}_{\rm D}$ we finally obtain:
\begin{equation}
\label{eq:HD_H2_local}
\frac{n_{\rm HD}}{2n_{\rm H_2}} = \cfrac{\rm D}{\rm H} \frac1{f_{\rm H_2}} \left(\cfrac{1}{\beta^{\rm chem} f_{\rm H_2} + \beta^{\rm dust}}+1\right)^{-1},
\end{equation}
where ${\rm D}/{\rm H} \equiv n_{\rm D}^{\rm tot} / n_{\rm H}^{\rm tot}$ is D to H isotopic ratio. One can write $\rm D /\rm H = \left({\rm D}/{\rm H}\right)_{\rm pr} a(Z)$, where $\left({\rm D}/{\rm H}\right)_{\rm pr}$ is the primordial value of the isotopic ratio, and $a(Z)$ is a factor of astration of D, which mainly is a function of metallicity. However, astration typically is small -- for the solar metallicity $a(Z_{\odot})\sim0.9$ \citep{Dvorkin2016}. A slightly different expressions for $n_{\rm HD}/n_{\rm H_2}$ ratio were written by e.g. \citealt{Federman1996, LePetit2002, Liszt2015}

Following \citealt{Sternberg2014}, $f_{\rm H_2}$ and $N_{\rm H}$ can be expressed as the analytical functions of $N_{\rm H_2}$ and specified by dimensionless parameter $\alpha$, which is the ratio of free space dissociation and formation rates of H$_2$
\begin{equation}
    \alpha \equiv \frac{\chi D^{\rm H2}}{R^{\rm H2} n^{\rm tot}_{\rm H}} = 1.3\times 10^{4} \frac{\chi}{n_{2}Z},
\end{equation}
where $D^{\rm H2}=5.8\times 10^{-11}$\,s$^{-1}$ is an unattenuated photodissociation rate of H$_2$ in Draine UV field and $R^{\rm H2} = 4.4 \times 10^{-17}$cm$^{3}$\,s$^{-1}$ is the formation rate of H$_2$ on the dust grains \citep{Sternberg2014}. Parameters $\beta^{\rm chem}$ and $\beta^{\rm dust}$ also depends on $N_{\rm H_2}$ and $N_{\rm HD}$. Therefore, we can use equation~\eqref{eq:HD_H2_local} to find $N_{\rm HD}$ as a function of $N_{\rm H_2}$. To do this we can substitute $n_{\rm HD}/n_{\rm H_2} = {\rm d} N_{\rm HD}/{\rm d} N_{\rm H_2}$ and solve the differential equation for $N_{\rm HD}(N_{\rm H_2})$ setting a boundary condition using $n_{\rm H_2}$ and $n_{\rm HD}$ values calculated in the unattenuated UV field.

Figure~\ref{fig:main} shows the calculated $N_{\rm HD}/2N_{\rm H_2}$ profiles as a function of $N_{\rm H_2}$ for fixed $n_2 = 1$, $\zeta_{\rm p} = 3$, $\chi = 1$ (corresponding to the typical conditions in ISM) and two different metallicities $Z=1$ (left panel) and $Z=0.1$ (right panel). In agreement with the previous studies \citep{LePetit2002, Liszt2015}, we find that the $N_{\rm HD}/2N_{\rm H_2}$ ratio significantly varies with a cloud depth and strongly depends on the ISM parameters. One can see, that at low metallicity HD/2H$_2$ ratio is significantly enhanced, and can be even higher than isotopic abundance.
This naturally explains the difference between measurement of HD/H$_2$ abundance at high z and in Milky Way \citep[see also,][]{Liszt2015}. The main driver for this enhancement is that the low metallicity favours the increase of ionization fraction of H$^{+}$ (and hence D$^{+}$), since the lower metallicity, the less destruction rate of H$^+$ appears in the right hand side of equation~\eqref{eq:CR_H}. 


\begin{figure*}
 \includegraphics[width=1.\textwidth]{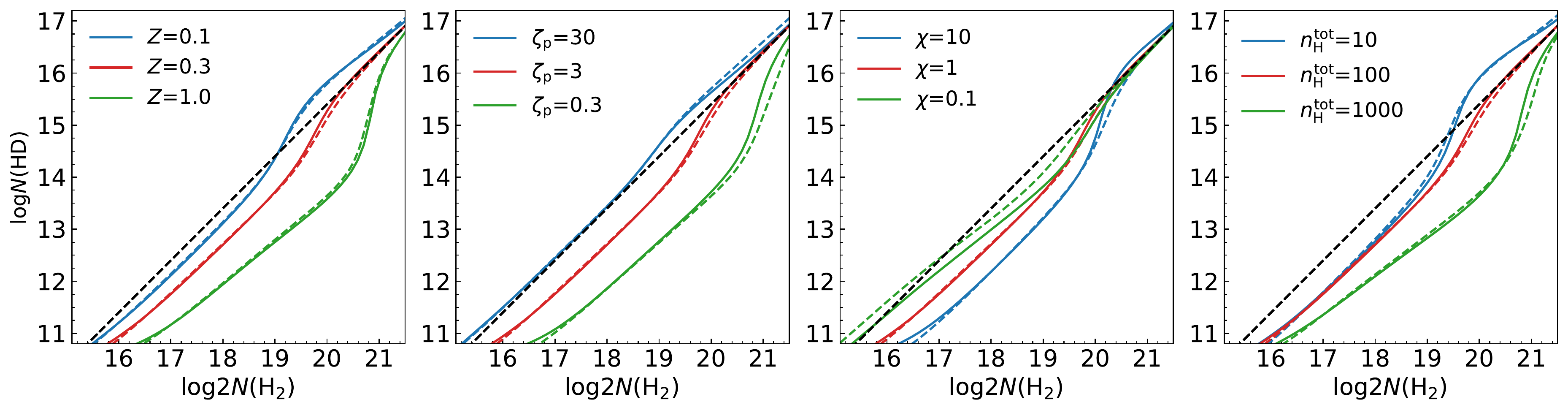}
 \caption{Dependence of the $N_{\rm HD}/N_{\rm H_2}$ on the variation of the physical parameters of the ISM. The red curves show the base model with $Z=0.3$, $\chi = 1$, $\zeta_{\rm p} = 3$ and $n_{\rm H}^{\rm tot} = 100$. Each panel provides the variation of one parameter from the base model: the metallicity, cosmic ray ionization rate, UV field strength and number density from left to right panel, respectively. The solid curves show solution of equation~\eqref{eq:HD_H2_local}, while the dashed curves are results from calculation by {\it Meudon PDR} code. The black dashed curve corresponds to the D/H isotopic ratio.
 }
\label{fig:HD_meudon}
\end{figure*}

\subsection{Asymptotics}
Analysing equation~\eqref{eq:HD_H2_local} we found three asymptotics of the $n_{\rm HD}/n_{\rm H_2}$ ratio. We note that these asymptotic values are almost the same as for the $N_{\rm HD}/N_{\rm H_2}$ ratio.

\begin{enumerate}
    \item Highly shielded region, $S^{\rm HD} \to 0$ and/or $\tau_b \gg 1$. 
    
    In that case $\beta^{\rm chem} \gg 1$, $\beta^{\rm dust} \gg 1$ and $f_{H_2} \to 1$. Therefore we get simply
    \begin{equation}
        \dfrac{n_{\rm HD}}{2n_{\rm H_2}} = \dfrac{\rm D}{\rm H},
    \end{equation}
    i.e. all H and D are in H$_2$ and HD molecules, respectively. However, this asymptotic works only for very high column density lines of sight (see Fig.~\ref{fig:main}), and in practice it is hard to observe, due to either high associated extinction or high H$_2$ column densities. Also we neglected other reactions, e.g. a destruction of H$_2$ and HD by cosmic rays, which can hinder the full molecularization in shielded parts of the clouds, the depletion on the dust grains and deuteration of complex molecules.
    
    \item Unattenuated UV field and intermediate $f_{\rm H_2}$.
    
    For $\log N_{\rm HD} \lesssim 14$ we have $S^{\rm HD} \to 1$ and $\tau_b \to 0$ for the reasonable values of ISM parameters. Depending on the H$_2$ molecular fraction, we obtained two asymptotic values. 
    For an intermediate molecular fraction, $f_{\rm H_2} \gtrsim 10^{-4}$,  
    $\beta_0^{\rm chem} f_{\rm H_2} > \beta_0^{\rm dust}$ and therefore
    \begin{equation}
    \label{eq:HD_H2_local_chem}
    \dfrac{n_{\rm HD}}{2n_{\rm H_2}} \approx \dfrac{\rm D}{\rm H} \beta_0^{\rm chem}, 
    \end{equation}
    where $\beta^{\rm chem}_0$ actually depends on $f_{\rm H_2}$ (see equation~\eqref{beta_0}), but for the typical ISM parameters it is close to the value for atomic hydrogen dominated region (see Fig.~\ref{fig:main}). 
    
    \item Unattenuated UV field and low $f_{\rm H_2}$.
    
    In the opposite case of very low H$_2$ molecular fraction (corresponds to the very surface of the medium, where we can write $f_{\rm H_2} = 4/\alpha$ following \citealt{Sternberg2014}) $\beta_0^{\rm dust}$ dominates and therefore we get
    \begin{equation}
    \label{eq:HD_H2_local_dust}
    \dfrac{n_{\rm HD}}{2n_{\rm H_2}} = \dfrac{\rm D}{\rm H} \dfrac{\beta_0^{\rm dust}}{f_{\rm H_2}} = \frac{\alpha}{4}\dfrac{\rm D}{\rm H} \beta_0^{\rm dust} = \dfrac{\rm D}{\rm H} \dfrac{R^{\rm HD} D^{\rm H_2}}{2 R^{\rm H_2} D^{\rm HD}} \approx 1.6 \dfrac{\rm D}{\rm H}.
    \end{equation}
    
    However the last asymptotic will not appear if $\beta_0^{\rm chem} \gtrsim 1.6$. Even more, it is actually not observable, since it is valid only at the unattenuated edge of the cloud, where the HD column density is very low. For the current instrumentation, a reasonable observational limit is $\log N_{\rm HD} \gtrsim 12.5$ (see Fig.~\ref{fig:main}). 
    
\end{enumerate}

\subsection{D\,{\sc i}/HD transition}
\label{sect:D_HD_transition}
The processes of HD formation and destruction imply that $n_{\rm HD}$ gradually increases with increasing penetration depth of UV radiation from some low value at the unattenuated edge to $n^{\rm tot}_{\rm D}$. Conversely, $n_{\rm D}$ is decreasing from $n^{\rm tot}_{\rm D}$, and hence there is a transition point between D and HD, which formally can be specified as $n_{\rm D} = n_{\rm HD}$. Using equation \eqref{eq_HD_2}, we can write the condition for the D\,{\sc i}/HD transition as
\begin{equation}
    \beta^{\rm chem} f_{\rm H_2}+ \beta^{\rm dust} \approx 1.
\end{equation}
    
For the reasonable range of ISM parameters, $\beta^{\rm chem} f_{\rm H_2} \gg \beta^{\rm dust}$ at D\,{\sc i}/HD transition. In other words, the chemical formation reaction \eqref{eq:H2+D+} determines D\,{\sc i}/HD transition and the $N_{\rm H_2}$ column density at which this transition occurs can be obtained from
\begin{equation}
    \label{eq:D_HD_tr_point}
   f_{\rm H_2} \approx 
   (\beta^{\rm chem}_0)^{-1} S^{\rm HD} e^{-\tau_g}. 
\end{equation}

Taking into account that $f_{\rm H_2} = 1/2$ formally determines the H\,{\sc i}/H$_2$ transition, one can see that D\,{\sc i}/HD transition will occur at lower penetration depth into the cloud than H\,{\sc i}/H$_2$ transition if $\beta^{\rm chem}_0 > 2 S^{\rm HD} e^{-\tau_{\rm tran}}$, where $\tau_{\rm tran}$ is the optical depth at which H\,{\sc i}/H$_2$ transition occurs, \citealt{Sternberg2016}. We find that such situation is typical for low metallicities (e.g. $Z\sim0.1$, see the right panel in Fig.~\ref{fig:main}), and it leads to $N_{\rm HD}/2N_{\rm H_2} > \rm D/H$ at the penetration depths around both D\,{\sc i}/HD and H\,{\sc i}/H$_2$ transitions.
 
\subsection{Comparison with the Meudon PDR code}
\label{sect:Meudon}
We used the {\it Meudon PDR} code \citep{LePetit2006} to check our calculation. We set a slab of the gas irradiated by the beamed radiation field (with the Drain spectrum) from one side and calculated several isochoric models with the fixed temperature $T = 100$\,K. The base model has the metallicity $Z=0.3$, cosmic ray ionization rate $\zeta_{\rm p} = 3$, hydrogen number density $n_{\rm tot}^{\rm H} = 100$\,cm$^{-3}$ and strength of UV field $\chi=1$. Then we varied independently $Z$, $\zeta_{\rm p}$, $n_{\rm tot}^{\rm H}$ and $\chi$ with values (0.1, 0.3, 1), (0.3, 3, 30), (10, 100, 1000) and (0.1, 1, 10), respectively. To better reproduce $\rm H_2$ profiles calculated in {\sl Meudon} we set the H$_2$ formation rate, $R_{\rm H_2}=8.0\times10^{-17}$\,cm$^{3}$s$^{-1}$. Although $N_{\rm HD}/N_{\rm H_2}$ significantly depends on the variation of each of these parameters, we found that our calculations are in reasonable agreement with {\it Meudon} results (see Fig.~\ref{fig:HD_meudon}).
The small discrepancy is due to difference in ionization fractions and slight difference in the HD and H$_2$ self-shielding functions.

\begin{figure}
    \centering
    \includegraphics[width=\columnwidth]{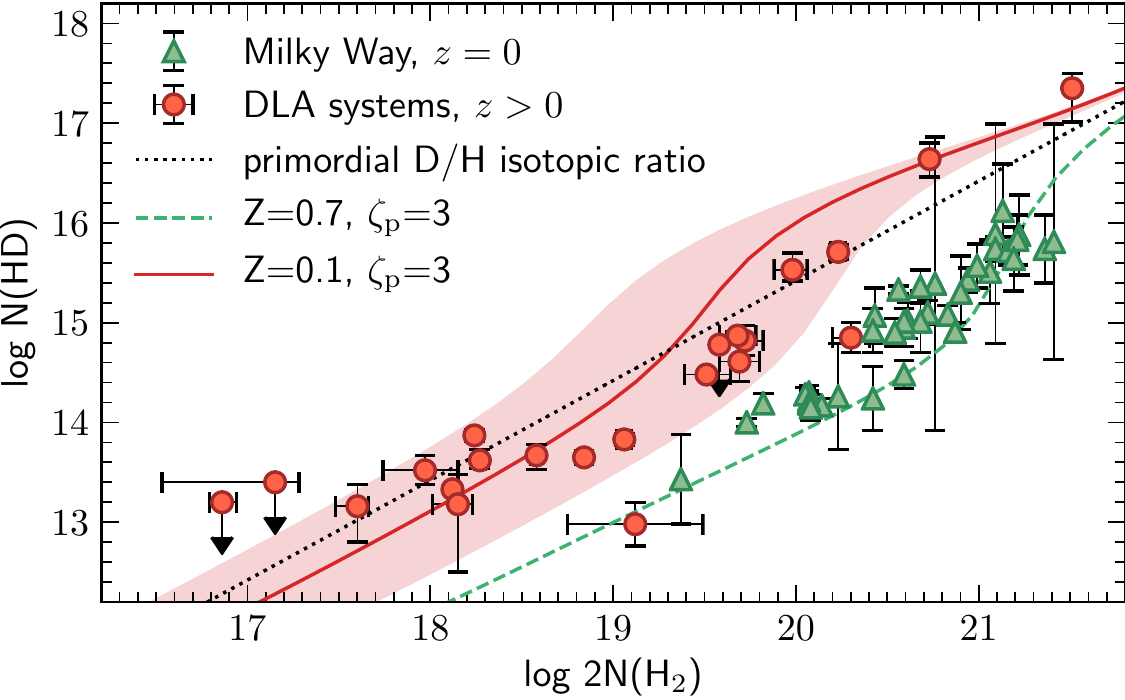}
    \caption{HD versus H$_2$ column densities. The red circles and green triangles show the measurements at high redshifts \citep[see ][and references therein]{Ivanchik2015, Kosenko2018} and in our Galaxy \citep{Snow2008}, respectively. The green dashed and red solid lines correspond to solution of equation~\eqref{eq:HD_H2_local} for $Z=0.7$ with  $\zeta_{\rm p}=3$, and $Z=0.1$ with $\zeta_{\rm p}=3$, respectively. Other parameters are $\chi = 1$, $n=100$\,cm$^{-3}$ and $T=100$\,K. The red filled stripe corresponds to variation of $\zeta_{\rm p}$ in range 0.3..30. The black dotted line indicates the D/H isotopic ratio.
    }
    \label{fig:HD_obs}
\end{figure}

\section{Discussion}
We showed that the HD/H$_2$ relative abundance is sensitive to the combination of physical conditions ($Z$, $\chi$, $n_{\rm tot}$ and $\zeta_{\rm p}$), and, therefore, it can be used as a complementary way to constrain them. Fortunately, in many cases the metallicity can be well constrained using the associated metal lines. Additionally, the number density and UV flux also can be estimated for some absorption systems from associated fine-structure levels of C\,{\sc i} and rotational levels of H$_2$ \citep{Balashev2019}. In Fig.~\ref{fig:HD_obs} we qualitatively show how our calculations can explain the measured $N_{\rm HD}/2N_{\rm H_2}$ in different environments. For illustrative purpose, we calculate HD/H$_2$ abundance for fixed $T=100$\,K, $n_{\rm H}^{\rm tot} = 100$\,cm$^{-3}$ and $\chi=1$. To compare with observations we used $Z=0.7$ for the Milky Way and depletion, $d=0.5$. One can see that the cosmic ray ionization rate $\zeta_{\rm p}\sim 3 \times 10^{-17}\,{\rm s^{-1}}$ agrees well with the local HD/H$_2$ measurements.
For comparison with DLA measurements we used $Z=0.1$ and $d=1$. Beside the fact that lower metallicity essentially explains the higher HD/H$_2$ relative abundance in DLAs than in Milky Way, the DLA measurements indicate much higher dispersion of $N_{\rm HD}/2N_{\rm H_2}$, than local ones. This probably reflects the large dispersion of the physical conditions in DLAs. Though in Fig.~\ref{fig:HD_obs} we show that the dispersion in DLA measurements can be reproduced using variation of $\zeta_{\rm p}$ between 30 and 0.3 (corresponding to the red shaded region), to accurately constrain $\zeta_{\rm p}$ we need to take into account the variation in other physical parameters. For example, we used value $Z=0.1$ that is close to the mean value obtained in high-z DLAs, however the dispersion is large.

We summarize that with a detailed analysis of absorption systems (both for local and high-z DLAs) $N_{\rm HD}/2N_{\rm H_2}$ ratio can be efficiently used to constrain physical conditions in the diffuse ISM. Out of which the cosmic ray ionization rate (or ionization fraction) is most important, since it is hard to constrain it by other means in the diffuse ISM.

\section*{Acknowledgements}

This paper was supported by RSF grant 18-12-00301. We thank an anonymous referee, Peter Shternin and Pasquier Noterdaeme the useful comments and suggestions.

\bibliographystyle{mnras}
\bibliography{references.bib}
\label{lastpage}

\end{document}